\documentclass[%
 reprint,
 superscriptaddress,
 amsmath,amssymb,
 aps,
 prc,
 nofootinbib
]{revtex4-1}

\usepackage[utf8]{inputenc} % allows certain fonts
\usepackage[colorlinks=true,allcolors=blue]{hyperref} % blue cross-refs
\usepackage[shortlabels]{enumitem} % Custom labels for lists
\usepackage[dvipsnames]{xcolor} % more colors

\usepackage{graphicx} % Include figure files
\usepackage{tabularx} % Tables evenly spaced
\usepackage{dcolumn}  % Align table columns on decimal point with column D
\newcolumntype{d}[1]{D{.}{.}{#1}} % Specify float format with column d{x.x}
\newcolumntype{R}{>{\raggedleft\arraybackslash}X} % Right aligned column in tabularx env
\usepackage{colortbl}
\definecolor{lgray}{gray}{0.9}
\newcolumntype{C}[1]{>{\centering\arraybackslash}p{#1}}

\usepackage{amsmath}
\usepackage{amsfonts}
\usepackage{xspace}
\usepackage{bm}
\usepackage{physics}

\renewcommand{\vec}[1]{\boldsymbol{#1}} % Vectors are bold

\newcommand{\pnfam}{\textsc{pnfam}}

\newcommand{\hfbtho}{\textsc{hfbtho}}

\begin{document}

%\preprint{APS/123-QED}

\title{Two-body weak currents in heavy nuclei}

\author{E. M. Ney}
\email[]{evan.ney@unc.edu}
\affiliation{Department of Physics and Astronomy, CB 3255, University of North
Carolina, Chapel Hill, North Carolina 27599-3255, USA\looseness=-1}

\author{J. Engel}
\email[]{engelj@unc.edu}
\affiliation{Department of Physics and Astronomy, CB 3255, University of North
Carolina, Chapel Hill, North Carolina 27599-3255, USA\looseness=-1}

\author{N. Schunck}
\email[]{schunck1@llnl.gov}
\affiliation{Nuclear and Chemical Science Division, LLNL, Livermore, California
94551, USA\looseness=-1}

\date{\today}

\begin{abstract}
In light and medium-mass nuclei, two-body weak currents from chiral effective
field theory account for a significant portion of the phenomenological quenching
of Gamow-Teller transition matrix elements.  Here we examine the systematic
effects of two-body axial currents on Gamow-Teller strength and $\beta$-decay
rates in heavy nuclei within energy-density functional theory.  Using a Skyrme
functional and the charge-changing finite amplitude method, we add the
contributions of two-body currents to the usual one-body linear response in the
Gamow-Teller channel, both exactly and though a density-matrix expansion.  The
two-body currents, as expected, usually quench both summed Gamow-Teller strength
and decay rates, but by an amount that decreases as the neutron excess grows.
In addition, they can enhance individual low-lying transitions, leading to decay
rates that are quite different from those that an energy-independent quenching
would produce, particularly in neutron-rich nuclei.  We show that both these
unexpected effects are related to changes in the total nucleon density as the
number of neutrons increases.
\end{abstract}
\maketitle

\section{\label{sec:intro}Introduction}

Beta decay is a well-studied weak process. Dating back to 1933, Fermi's theory
of $\beta$ decay~\cite{Stuewer1995} paved the way for our later understanding of
the electro-weak force. Despite substantial progress, however, a peculiar
feature of nuclear $\beta$ decay has puzzled physicists for decades.
Gamow-Teller transition rates, the primary contributions to decay in most
nuclei, are systematically over-predicted by the nuclear shell
model~\cite{Brown1988}, and a phenomenological ``quenching factor'' has been
required to bring theoretical predictions in line with experimental
data~\cite{Chou1993,MartinezPinedo1996,Kumar2016}.  The
physical source of the quenching was unclear until fairly recently.

The literature contains several reviews of the so-called quenching problem, recent examples
include Refs.~\cite{Suhonen2017, Engel2017}. The work of
Ref.~\cite{Gysbers2019} provided compelling evidence that a significant portion
of the quenching comes from two sources: nuclear correlations and two-body
meson-exchange currents.  Nucleons contribute coherently to the two-body
currents, which should therefore be important in heavy nuclei. Although
\textit{ab initio} many-body methods, with interactions and currents from chiral
effective field theory ($\chi$EFT), have proved useful for studying the
quenching problem in lighter nuclei, they are computationally difficult to apply
in most heavy nuclei. Several studies of double-$\beta$ decay and dark-matter
scattering in medium-mass and heavy nuclei treated two-body currents in a simple
nuclear-matter approximation
\cite{Menendez2011,Menendez2012,Klos2013,Engel2014}, but a more complete
treatment in heavy systems is still missing. Here we fill the gap.

Our approach is to use the $\chi$EFT currents in conjunction with a Skyrme
energy functional.  The simultaneous use of two distinct schemes is
inconsistent, of course, but will do for an initial investigation of the effects
of the two-body currents.  Ultimately, we want to treat the numerical
coefficients of the chiral currents as parameters to be fit in conjunction with
the energy-density functional. We defer that large task to a future paper.

This paper is structured as follows: In Sec.~\ref{sec:theory} we discuss the
weak axial current, its implementation in nuclear energy-density-functional
(EDF) calculations of linear response, and a density matrix expansion of the
two-body current.  In Sec.~\ref{sec:method}, we outline our method for
numerically incorporating the two-body current in an axially-deformed oscillator
basis. We then present calculations in several nuclei and discuss the effects of
the new current in Sec.~\ref{sec:results}.  Finally, in
Sec.~\ref{sec:conclusions}, we discuss the outlook and conclude.

\section{\label{sec:theory}Theoretical Model}
\subsection{\label{ssec:currents}Weak axial current}
Beta decay is a semi-leptonic process. It is governed by a Hamiltonian density
that for energies much less than the mass of the $W$ boson can be written as
\begin{equation}
\label{eq:h_beta}
    {H}_\beta = -\frac{G_\beta}{\sqrt{2}} \int d^4 x\ {J}_{\mu}(x)
    {L}^\mu(x) + \text{h.c.} 
    \,,
\end{equation}
where $G_\beta/(\hbar c)^3 =1.14959 \times 10^{-5}$ GeV$^{-2}$ can be inferred
from super-allowed decays~\cite{Hardy2020}, ${J}_{\mu}(x)$ is the nuclear weak
current, ${L}^\mu(x)$ is the leptonic weak current, $x\equiv(\vec{x},t)$, and we
use the Einstein sum convention.  With Fermi's golden rule in first-order
perturbation theory, one computes decay rates from a phase-space-weighted
transition matrix element.  Equation~\eqref{eq:h_beta} implies that the $\beta$-decay
transition matrix element between the initial atomic state $\ket{I}$ and the
final state $\ket{F}$ is \cite{Walecka1975}
\begin{equation}
\label{eq:hbeta_mtxel}
\bra{F} {H_\beta} \ket{I} 
= -\frac{G_\beta}{\sqrt{2}} \, l^\mu 
\int \dd^3x\ \bra{f} e^{-i\vec{q}\cdot \vec{x}} {J}_{\mu}(\vec{x}) \ket{i}
\,.
\end{equation}
In this expression $\vec{q}$ is the momentum transferred to leptons and $l^\mu$
is a leptonic matrix element that depends only on initial and final lepton wave
functions.  The remaining term, $\bra{f} {J}_{\mu}(\vec{x}) \ket{i}$, is the
matrix element of the nuclear current operator between initial and final nuclear
states $\ket{i}$ and $\ket{f}$.  Because nuclear states are complicated, the
nuclear matrix elements cannot be written in closed form.  Most often, one
invokes the impulse approximation, which takes the nucleus to be a collection of
free nucleons, so that the current is represented by a one-nucleon operator. 

In the Standard Model, the nuclear current is a sum of vector and axial-vector
pieces, $J_\mu(\vec{x}) = J^V_\mu(\vec{x}) + J^A_\mu(\vec{x})$.  Here, we
consider only the axial current in the limit of zero momentum transfer. The
resulting leading-order contributions come from the spatial piece of the
four-current.  In the non-relativistic impulse approximation this piece is a is
a one-body vector operator with the first-quantized form,
\begin{equation}
\label{eq:1bc}
\vec{J}^A_{\text{1b}}(\vec{x}) = -g_A \sum_i \vec{\sigma}^{(i)}
t^{(i)}_{\pm}   \delta^3(\vec{x} - \vec{r}_i) \,.
\end{equation}
The sum is over nucleons in the nucleus, $g_A \approx 1.27$~\cite{PDG2020} is
the axial-vector coupling, $\vec{x}$ is the position --- a $c$-number argument
of the quantum field operator $\vec{J}^A_{\text{1b}}$ --- and $\vec{r}_i$ is an
operator describing the location of the $i^{\mathrm{th}}$ nucleon relative to
the nuclear center of mass.  We use the notation $\vec{\sigma}^{(i)} \equiv
(\sigma_x^{(i)},\sigma_y^{(i)},\sigma_z^{(i)})$ and $\vec{\tau}^{(i)} \equiv
(\tau_x^{(i)},\tau_y^{(i)},\tau_z^{(i})$ for the usual Pauli matrices acting on
the two-component spin and isospin vectors of the $i^{\text{th}}$ nucleon.  We
use $\vec{t}$ to denote the isospin operators themselves: $\vec{t}^{(i)}
=\tfrac{1}{2}\vec{\tau}^{(i)}$. The raising/lowering operators in
Eq.~\eqref{eq:1bc_op_fam} are then $t_{\pm}^{(i)} = t_x^{(i)} \pm i t_y^{(i)} $.
With this notation, $t_- \ket{n} = \ket{p}$, $t_+ \ket{p} = \ket{n}$, and $t_-
\ket{p} = t_+\ket{n}=0$.  

To go beyond the impulse approximation, we need to specify degrees of freedom.
We take them to be given by $\chi$EFT, which treats only nucleons and pions
explicitly, including other effects in contact interactions among the
constituents.  We restrict ourselves to the leading-order two-body axial current
derived in Eqs.~A5 and A6 of Ref.~\cite{Park2003}. From those expressions, we
derive the first-quantized two-body current operator, which has the form,
\begin{equation}
\label{eq:2bc}
\begin{aligned}
    \vec{J}^A_{\mathrm{2b}}(\vec{x}) = \sum_{i<j}
    \Big[
    \mathcal{O}^{\mathrm{(2b)}}_{ij}&(\vec{r}_i - \vec{r}_j, \phantom{-} \vec{p}_{i})\ \delta(\vec{x} - {\vec{r}}_i) 
    \\&
    + 
    \mathcal{O}^{\mathrm{(2b)}}_{ji}(\vec{r}_j - \vec{r}_i, -\vec{p}_j)\ \delta(\vec{x} - {\vec{r}}_j)
    \Big]
    \,.
\end{aligned}
\end{equation}
We leave the operators $\mathcal{O}^{\mathrm{(2b)}}$ unspecified for now, using
them here just to indicate the form of the current.  They depend on the
coordinates of two nucleons, and the momentum $\vec{p}_i = -\tfrac{i}{2}
(\overrightarrow{\grad}_i - \overleftarrow{\grad}_i)$, which acts on wave
functions of the $i^{\mathrm{th}}$ nucleon.

We use the axial current operators in conjunction with
Eq.~\eqref{eq:hbeta_mtxel} to generate a set of $\vec{x}$-independent operators
whose nuclear transition matrix elements determine the decay probability. We
begin by substituting Eqs.\ \eqref{eq:1bc} and \eqref{eq:2bc} in Eq.\
\eqref{eq:hbeta_mtxel}, and then evaluate the integral over $\vec{x}$. Using the
long-wavelength (or allowed) approximation, we assume that $\abs{\vec{q}} R$,
where $R$ is the nuclear radius, is small enough to let us replace $e^{- i
\vec{q} \cdot \vec{x}}$ by unity.  For the one-body part of the current, the
procedure generates the operator 
\begin{equation}
\label{eq:1bc_op_fam}
{\vec{A}}_{\text{1b}}  = -g_A \sum_i\vec{\sigma}^{(i)} t^{(i)}_{\pm} \,.
\end{equation}
The two-body current contains a short-range (contact) piece and a finite-range
pion-exchange piece.  For the short-range part, after inserting explicit
expressions for the operators $\mathcal{O}^{(\mathrm{2b})}$ in
Eq.~\eqref{eq:2bc}, we obtain, 
\begin{multline}
\label{eq:2bc_op_fam_delt}
\vec{A}_{\text{2b},s}= -\frac{g_A}{4 m_N f_{\pi}^2} \sum_{i<j} \Big[
2\bar{d}_1 \big( t^{(i)}_{\pm} \vec{\sigma}^{(i)} + t^{(j)}_{\pm}
\vec{\sigma}^{(j)} \big) \\
+ 4 \bar{d}_2\, t^{(i\times j)}_{\pm} \vec{\sigma}^{(i\times j)} \Big]
\delta^3(\vec{r}_i-\vec{r}_j) \,,
\end{multline}
where $s$ stands for ``short,'' $\vec{\sigma}^{(i\times j)} \equiv
\vec{\sigma}^{(i)} \times \vec{\sigma}^{(j)}$, $t^{(i\times j)}_\pm \equiv
(\vec{t}^{(i)} \times \vec{t}^{(j)})_x \pm i (\vec{t}^{(i)} \times
\vec{t}^{(j)})_y$, and to arrive at this expression we have inserted a missing factor of
$-\frac{1}{4}$~\cite{Krebs2017,Wang2018,Gysbers2019} in the short-range term
from Eq.~A6 of Ref.~\cite{Park2003}. The
finite-range pion-exchange two-body operator is
\begin{widetext}
\begin{multline}
\label{eq:2bc_op_fam_y0}
\vec{A}_{\text{2b},\pi} = - \frac{g_A}{2 m_N f_{\pi}^2} \bigg\{ 4\bar{c}_3
\sum_{i<j} \Big[ t_{\pm}^{(j)} \big(\vec{\sigma}^{(j)}\cdot\grad\big)\grad +
t_{\pm}^{(i)} \big(\vec{\sigma}^{(i)}\cdot\grad\big)\grad \Big] + 2
t_{\pm}^{(i\times j)} \Big[ {\vec{p}_{i}} \big(\vec{\sigma}^{(j)} \cdot
\grad\big) + {\vec{p}_{j}} \big(\vec{\sigma}^{(i)} \cdot \grad\big) \Big] \\
+ 4\left(\bar{c}_4 + \frac{1}{4}\right) t_{\pm}^{(i\times j)} \Big[
\big(\vec{\sigma}^{(i)} \times \grad\big) \big(\vec{\sigma}^{(j)} \cdot
\grad\big) - \big(\vec{\sigma}^{(j)} \times \grad\big) \big(\vec{\sigma}^{(i)}
\cdot \grad\big) \Big] \bigg\} Y_0(\abs{\vec{r}_i -\vec{r}_j}) \,.
\end{multline}
\end{widetext}
This expression is equivalent to those given in Refs.~\cite{Park2003,Wang2018},
but is written entirely in terms of derivatives acting on the Yukawa function,
$Y_0(r) = e^{-m_\pi r}/(4 \pi r)$.  The dimensionless low energy constants
(LECs) in Eqs.~\eqref{eq:2bc_op_fam_delt}-\eqref{eq:2bc_op_fam_y0} are defined
by \cite{Park2003}
\begin{equation}
\label{eq:lecs}
\bar{c}_i = m_N c_i, \qquad \bar{d}_i = \frac{m_N f_\pi^2}{g_A} d_i 
\,.
\end{equation}
% We take the nucleon mass to be $m_N=939$~MeV, the pion decay constant to be
% $f_\pi = 92.4$~MeV, and the pion mass to be $m_\pi=138.04$~MeV.

\subsection{\label{ssec:qrpa}Linear response in energy-density-functional theory}

Nuclear EDF theory away from closed shells is implemented through the
Hartree-Fock-Bogoliubov (HFB) generalized mean field.  To the extent that static
HFB calculations approximate the exact energy and one-body density, the
time-dependent HFB equations provide an adiabatic approximation to the full time
evolution of the nuclear density, and the HFB linear response is an adiabatic
approximation to the exact response.  The time-dependent equations can be
written as \cite{Ring2004}
\begin{equation}
\label{eq:tdhfb}
    i \dot{R}(t)
    = 
    [\mathcal{H} + \mathcal{F}(t), R(t)]
    ,\quad
    R = \begin{pmatrix} \rho &  \kappa \\ -\kappa^* & 1-\rho* \end{pmatrix} 
    \,,
\end{equation}
with $\rho$ the one-body density matrix and $\kappa$ the pairing tensor. In
these equations, the mean-field Hamiltonian $\mathcal{H}$ and external field
$\mathcal{F}$ take the form
\begin{equation}
\begin{gathered}
\label{eq:tdhfb_HandF}
    \mathcal{H} = \begin{pmatrix} t + \Gamma & \Delta \\ -\Delta^* & -(t + \Gamma)^* \end{pmatrix}
    ,\ 
    \mathcal{F} = \begin{pmatrix} f + \widetilde{\Gamma} & \widetilde{\Delta} \\ -\widetilde{\Delta}^* &
    -(f + \widetilde{\Gamma})^* \end{pmatrix}
    \,.  
\end{gathered}
\end{equation}
The matrices $t$ and $f$ are from the kinetic piece of the Hamiltonian and the
one-body external field, respectively. The terms $\Gamma$ and $\Delta$ represent
the effects of correlations.  If the functional is associated with a (typically
density-dependent) potential $\hat{V}$, and if the external field is generated
in part, as it will be here, by a two-body operator $\hat{F} (t)$, these terms
come from contracting the two-body matrix elements of $\hat{V}$ and $\hat{F}
(t)$ with one-body density matrices:
\begin{equation}
\label{eq:mean_fields}
\begin{aligned}
\Gamma_{ik} &= \sum_{jl} \bar{v}_{ijkl}\rho_{lj}&
,\quad
\Delta_{ij} &= \frac{1}{2} \sum_{ij} \bar{v}_{ijkl}\kappa_{kl}\\
\widetilde{\Gamma}_{ik} &= \sum_{jl} \bar{F}_{ijkl}\rho_{lj}&
,\quad
\widetilde{\Delta}_{ij} &= \frac{1}{2} \sum_{ij} \bar{F}_{ijkl}\kappa_{kl}
\,,
\end{aligned}
\end{equation}
where the $\bar{v}$'s are antisymmetric matrix elements of $\hat{V}$ and the
$\bar{F}$'s are antisymmetric matrix elements of $\hat{F} (t)$.

The HFB linear response comes from treating the time-dependent HFB equations in
first order in the external field $\mathcal{F}(t)$ to obtain small oscillations
around the static mean field.  The oscillations are the same as those imposed by
the quasiparticle random phase approximation (QRPA). One can obtain the linear
response efficiently with the finite amplitude method
(FAM)~\cite{,Nakatsukasa2007,Avogadro2011}, the charge-changing version of which
is outlined in several places~\cite{Mustonen2014,Shafer2016,Ney2020}.  Our work
here takes advantage of the similarity between $\Gamma$ and $\tilde{\Gamma}$ in
Eq.\ \eqref{eq:mean_fields} to include the effects of the two-body nuclear
current operator in Eqs.~\eqref{eq:2bc_op_fam_delt}-\eqref{eq:2bc_op_fam_y0}. We
neglect the ``pairing field'' $\widetilde{\Delta}$ in this paper and compute
only the ``particle-hole mean field'' $\widetilde{\Gamma}$, which should be more
important.

Our computation involves external perturbations that change the projection along
the symmetry axis of the angular momentum by an amount $K$. In deformed nuclei
each component of the vector operators in
Eqs.~\eqref{eq:1bc_op_fam}-\eqref{eq:2bc_op_fam_y0} can induce a different
response, but in systems where time-reversal symmetry is conserved the response
to operators that change angular momentum by $\pm K$ is the same. The lab-frame
response is a linear combination of the intrinsic responses to the different
components of the vector operators, where each intrinsic result is multiplied by
the factor $\Theta_K = 1 \delta_{K,0} + \sqrt{2} \delta_{K,K>0}$
\cite{Bohr1998}. Because we therefore must compute the response to each component of
the current operators, we refer to the external field as a vector,
$\vec{\mathcal{F}}$, comprising $\vec{f}$, $\widetilde{\vec{\Gamma}}$, and
$\widetilde{\vec{\Delta}}$.  Equations \eqref{eq:tdhfb}-\eqref{eq:mean_fields}
should be understood to represent the response to a single component of these
vectors that changes intrinsic angular momentum by $K=0,\pm1$.

\section{\label{sec:method}Computational method}

We use the Python program PyNFAM~\cite{Ney2020} to compute $\beta$-decay
properties. This program wraps the ground-state HFB solver
{\hfbtho}~\cite{Stoitsov2005,Stoitsov2013,NavarroPerez2017} and the
charge-changing FAM solver {\pnfam}~\cite{Mustonen2014} for computing the 
linear response. In all our calculations we use the same SKO' Skyrme functional 
used in the global $\beta$-decay calculations of Refs.~\cite{Mustonen2016,Ney2020}. In
fitting this functional, the authors took an effective value for $g_A$ of 1.0,
and considered one-body Gamow-Teller strength only. 

Because of the similarity of Eqs.~\eqref{eq:tdhfb} and~\eqref{eq:tdhfb_HandF} to
static HFB equations, we can implement the full two-body current in the FAM in a
way that is very similar to the implementation of the finite-range Gogny
interaction in HFB calculations ~\cite{NavarroPerez2017}. This approach entails
computing two-body matrix elements of the finite-range operator and contracting
them with the one-body density, as in Eq.~\eqref{eq:mean_fields}. Unlike the
Gogny interaction, however, the two-body axial current is a charge-changing
spin-dependent vector operator with a finite-range part that involves a Yukawa
function (rather than a Gaussian).

\subsection{\label{ssec:contact_term}Contact term}

The zero-range term of the two-body current can be treated just like a Skyrme
interaction. We express the contraction of the antisymmetrized current operator
with the density matrix as
\begin{equation}
\widetilde{\vec{\Gamma}}_{ij, s} = \sum_{kl} \bra{ik}\bar{\vec{A}}_{2b,s}\ket{jl} \rho_{lk} ,
\label{eq:2b_contact_contract}
\end{equation}
where $\bar{\vec{A}}_{2b,s} = \vec{A}_{2b,s}(1 - P^r P^\sigma P^\tau)$ is the
antisymmetrized version of Eq.~\eqref{eq:2bc_op_fam_delt} and the matrix
elements are given by
\begin{multline}
\bra{ik}\bar{\vec{A}}_{2b,s}\ket{jl} = \int d^3 \vec{r} \sum_{\sigma_1 \sigma_1'
\tau_1 \tau_1'} \sum_{\sigma_2 \sigma_2' \tau_2 \tau_2'}\\
\times \bra{\vec{r} \sigma_1 \tau_1 \vec{r} \sigma_2 \tau_2}
\bar{\vec{A}}_{2b,s} \ket{\vec{r} \sigma_1' \tau_1' \vec{r} \sigma_2' \tau_2'}
\,. 
\end{multline}
We can rewrite Eq.~\eqref{eq:2b_contact_contract} in terms of the non-local
densities~\cite{Schunck2019,Perlinska2004},
\begin{equation}
\label{eq:nonlocal_densities}
\begin{aligned}
\rho_{00}(\vec{r},\vec{r}') 
&= \sum_{ij} \rho_{ji} \sum_{\sigma\tau} \phi^*_i(\vec{r}' \sigma \tau) \phi_j(\vec{r}\sigma \tau) ,
\\
\rho_{1k}(\vec{r},\vec{r}') 
&= \sum_{ij} \rho_{ji} \sum_{\sigma\tau\tau'} \phi^*_i(\vec{r}' \sigma \tau') \phi_j(\vec{r}\sigma \tau) \bra{\tau}\tau_k \ket{\tau'} ,
\\
\vec{s}_{00}(\vec{r},\vec{r}') 
&= \sum_{ij} \rho_{ji} \sum_{\sigma\sigma'\tau} \phi^*_i(\vec{r}' \sigma' \tau) \phi_j(\vec{r}\sigma \tau) \bra{\sigma}\vec{\sigma}\ket{\sigma'} ,
\\
\vec{s}_{1k}(\vec{r},\vec{r}') 
&= \sum_{ij} \rho_{ji} \sum_{\sigma\sigma'\tau\tau'} \phi^*_i(\vec{r}' \sigma' \tau') \phi_j(\vec{r}\sigma \tau) \\
& \qquad\qquad\qquad\times \bra{\sigma}\vec{\sigma}\ket{\sigma'} \bra{\tau}\tau_k \ket{\tau'}
\, ,
\end{aligned}
\end{equation}
which are, respectively, the scalar-isoscalar, scalar-isovector,
vector-isoscalar and vector-isovector components of the full one-body density
matrix. Then, evaluating the matrix elements of $\vec{A}_{\mathrm{2b},s}$, we
can extract the direct mean field current
$\widetilde{\vec{\Gamma}}_{s}^\text{dir.} (\vec{r})$, 
\begin{multline}
\label{eq:2b_contact_dir}
\widetilde{\vec{\Gamma}}_{s}^\text{dir.}(\vec{r}) = \frac{g_A}{2 m_N f_{\pi}^2}
\left\{ -\bar{d}_1 \left[ \rho_{00}(\vec{r}) \vec{\sigma} t_{\pm } \mp
\frac{1}{\sqrt{2}}\vec{s}_{1\pm 1}(\vec{r}) \right] \right.  \\
\left. - 2 i \bar{d}_2 \vec{\sigma} \times \left[ \vec{s}_{10}(\vec{r}) t_{\pm}
\pm \frac{1}{\sqrt{2}} \vec{s}_{1\pm 1}(\vec{r}) t_z \right] \right\} \,,
\end{multline}
where the densities with one coordinate $\vec{r}$ are the diagonal elements of
those defined in Eq.~\eqref{eq:nonlocal_densities}, and we have left the spin and
isospin components of the field in operator form.  When we compute matrix
elements, we get
\begin{equation}
\begin{aligned}
\widetilde{\vec{\Gamma}}^{\text{dir.}}_{ij,s} 
& = \bra{i}\widetilde{\vec{\Gamma}}_{s}^{\text{dir.}}(\vec{r}) \ket{j} \\
& = \int d^3\vec{r} \sum_{\sigma\tau,\sigma'\tau'} 
\phi_{i}^{*}(\vec{r}\sigma\tau)\widetilde{\vec{\Gamma}}_{s}^{\text{dir.}}(\vec{r})\phi_{j}(\vec{r}\sigma'\tau')
\,.
\end{aligned}
\end{equation}

To obtain the exchange part of the mean field, we evaluate the matrix elements
of $-\vec{A}_{2b,s} P^r P^\sigma P^\tau$ and extract the field 
\begin{multline}
\label{eq:2b_contact_exc}
\widetilde{\vec{\Gamma}}_{s}^\text{exc.}(\vec{r}) \\
= \frac{g_A}{2 m_N f_{\pi}^2} \left\{ \frac{1}{2}(\bar{d}_1 - 2\bar{d}_2) \left[
\rho_{00}(\vec{r}) \vec{\sigma} t_{\pm} \mp \frac{1}{\sqrt{2}} \vec{s}_{1\pm
1}(\vec{r}) \right] \right.  \\
\left.  + \frac{1}{2}(\bar{d}_1 + 2\bar{d}_2) \left[ \mp \frac{1}{\sqrt{2}}
\vec{\sigma} \rho_{1,\pm1}(\vec{r}) + \vec{s}_{00}(\vec{r}) t_{\pm} \right]
\right.  \\ 
\left.  - i\bar{d}_1 \left( \vec{\sigma} \times \left[ \vec{s}_{10}(\vec{r})
t_{\pm} \pm \frac{1}{\sqrt{2}} \vec{s}_{1\pm 1}(\vec{r})t_z \right] \right)
\right\} \,.
\end{multline}

Finally, our HFB ground state is symmetric under time-reversal and does not
include proton-neutron mixing. The spin and charge-changing ground-state
densities therefore vanish, leading to the result 
\begin{equation}
\label{eq:contact_term}
\begin{aligned}
\widetilde{\vec{\Gamma}}_{s}(\vec{r}) & \equiv \widetilde{\vec{\Gamma}}_{s}^\text{dir.}
(\vec{r}) +
\widetilde{\vec{\Gamma}}_{s}^\text{exc.}(\vec{r}) \\
    &=
    -\frac{g_A}{4 m_N f_\pi^2}  \bar{c}_D\, \rho_{00}(\vec{r})\, \vec{\sigma} t_{\pm}
    \,.
\end{aligned}
\end{equation}
Equation \eqref{eq:contact_term} is just as easy to work with as the usual
one-body Gamow-Teller operator and depends on a single combination $\bar{c}_D
\equiv \bar{d}_1 +2 \bar{d}_2$ of the LECs.

\subsection{\label{ssec:finite_range}Finite-range term}

To obtain the finite-range contribution to the mean field we must contract the
antisymmetrized finite-range part of the current with the density matrix
according to Eq.~\eqref{eq:mean_fields}. The two-body current contains Yukawa 
functions, which are not separable. Without separability, the time it would 
take to compute the mean field in a model space with $N$ harmonic oscillator 
shells would scale like $\mathcal{O}(N^{12})$, making large model spaces 
impossible \cite{Parrish2013}. 

However, we can approximate the Yukawa function by a sum of Gaussians. The two
functions do not have the same behavior at $r=0$ but the integrands in their
matrix elements, which contain a factor of $r^2$, do. We therefore use the
fit~\cite{Dobaczewski2009},
   \begin{equation}
   \label{eq:yukawa_approx}
   \begin{aligned}
      r^2 \bigg[ \frac{e^{-r}}{r} \bigg] \approx r^2 \bigg[ &6.79 e^{-34 r^2} +
      0.786 e^{-1.44 r^2} + 0.241 e^{-0.38 r^2} \\&
      - 0.062 e^{-0.15 r^2} + 0.078 e^{-0.13 r^2} \bigg] \,,
   \end{aligned}
   \end{equation}
to approximate two-body Yukawa matrix elements. The separability of Gaussian
interactions makes contraction with the density in configuration space
tractable. Because we perform our calculations in a basis of axially-deformed
harmonic oscillator states, in which the mean field contains a Cartesian ($z$)
component, contributing a computation time of $\mathcal{O}(N^{8})$, and a radial
($r,\phi$) component, contributing a time of $\mathcal{O}(N^{10})$, the
separability provides vastly better computational scaling. Thus, we work with
the non-local mean field $\widetilde{\vec{\Gamma}}_\pi (\vec{r}\sigma\tau,
\vec{r}'\sigma' \tau')$.  Spin and isospin degrees of freedom, however, are
implicit in {\hfbtho} and {\pnfam}. It is therefore necessary to sum over the
corresponding quantum numbers analytically. The result is then contracted over
spatial quantum numbers numerically.

We compute Gaussian matrix
elements in our axially-deformed oscillator basis in the way described in
Refs.~\cite{Younes2009,NavarroPerez2017}, obtaining them as analytic functions
of the oscillator quantum numbers, and using the Cartesian expansion of the
radial oscillator wave functions to express the full Gaussian matrix elements
entirely in terms of one-dimensional Gaussian matrix elements (i.e., with
one-dimensional wave functions).  The decomposition makes it easy to compute the
derivatives in the two-body current.  The finite-range terms in the current
appear not as the Yukawa function $Y_0(r)$, but in the form $\partial_i
\partial_j Y_0(r)$ and $(\overleftarrow{\grad}-\overrightarrow{\grad}) 
\partial_j Y_0(r)$.  The derivatives acting on $Y_0(r)$ can be integrated by 
parts so that they act on wave functions, allowing us to use relations that 
relate derivatives of one-dimensional oscillator wave functions to linear combinations of a few 
other such wave functions. Thus, all finite-range matrix elements in $\widetilde{\vec{\Gamma}}_\pi$ are expressed as linear combinations of one-dimensional Gaussian matrix elements.

From now on we neglect the terms proportional to $\vec{p}_{1}$ and $\vec{p}_{2}$
in Eq.~\eqref{eq:2bc_op_fam_y0}. They are computationally expensive to evaluate
and contribute very little. In the nuclear-matter approximation of
Refs.~\cite{Klos2013,Menendez2011}, for example, these terms change matrix
elements by only $1.5\%$--$2.5\%$ in the zero-momentum limit for typical values
of ${2 \bar{c}_4 - \bar{c}_3}$. In addition, time-reversal symmetry causes the
contributions of direct terms proportional to $\bar{c}_4$ to vanish exactly.

\subsection{\label{ssec:dme}Density matrix expansion}

To validate our implementation of the linear response produced by the two-body
current, and to find good approximations to it, we compare it to the response
produced by an effective density-dependent one-body current, derived from a kind
of density matrix expansion (DME).  The DME is a method to construct a
density-dependent, {\it local} operator $O(\vec{r})$ that approximates the
non-local operator $O(\vec{r},\vec{r}')$ of interest
\cite{Negele1972,Gebremariam2010,Bogner2011,Dyhdalo2017, NavarroPerez2018}. When
applied to a local, two-body, finite-range potential of the type
$V(\vec{r}_1,\vec{r}_2)$, the DME effectively maps it into a local one-body
potential $V(\vec{R})$ with $\vec{R} = \tfrac{1}{2}(\vec{r}_1 + \vec{r}_2)$.
The application to a charge-changing current operator rather than a
charge-conserving Hamiltonian introduces some subtleties that we describe along
with details of the expansion in Appendix~\ref{app:dme}. The DME leads to a more
sophisticated density dependence for the one-body current than that given in
Refs.~\cite{Menendez2011} and \cite{Menendez2012}.

At leading order, the DME reproduces the contact term exactly (cf.
Eq.~\eqref{eq:contact_term}). As for the finite-range piece, the expansion
produces no direct term at all at leading order. There is a leading-order
exchange term, however, given by 
\begin{multline}
\label{eq:DME_exchange}
\widetilde{\vec{\Gamma}}_\text{DME}^{\text{exc.}} ( \vec{r}) = \frac{g_A}{m_N
f_\pi^2} \vec{\sigma} t_\pm \\
\times \bigg\{ \bar{C} \bigg[ 1 - m_\pi^2 F\left[ k_F(\vec{r}) \right] -
\frac{1}{10}m_\pi^2 G\left[ k_F(\vec{r}) \right] \bigg]\rho (\vec{r}) \\
- \bar{D} \rho(\vec{r}) - \frac{1}{6}\frac{\bar{C} m_\pi^2}{k_F^2(\vec{r})}
G\left[ k_F(\vec{r}) \right] \bigg[ \frac{1}{4}\nabla^2\rho(\vec{r}) -
\tau(\vec{r})\bigg] \bigg\} \,,
\end{multline}
with $\bar{C} = \frac{1}{3}\left( 2 \bar{c}_4 - \bar{c}_3 + \frac{1}{2}\right)$
and $\bar{D} \equiv \bar{d}_1 - 2 \bar{d}_2$. Here $\rho (\vec{r})$ is the
scalar-isoscalar particle density, $\rho (\vec{r}) \equiv \rho_{00}(\vec{r})$,
and $\tau(\vec{r})$ is the (isoscalar) kinetic density, ${\tau(\vec{r}) \equiv
\grad \cdot \grad' \rho_{00}(\vec{r},\vec{r}') \rvert_{\vec{r}=\vec{r}'}}$. The
functions $F$ and $G$ are given by
\begin{equation}
\label{eq:DME_U}
\begin{aligned}
    F[k] &= \frac{3}{2k^2} \bigg( 1 - \frac{m_\pi}{k} \tan^{-1}\left(\frac{2k}{m_\pi}\right)
     \\ &\qquad\qquad\qquad
     + \frac{m_\pi^2}{4 k^2} \ln\left( 1 + 4 \frac{k^2}{m_\pi^2}\right) \bigg)
    \\    
    G[k] &= \frac{3}{2k^2} \bigg( 1 + 8\frac{k^2}{4k^2 + m_\pi^2} 
     \\ &\qquad\qquad\qquad
    - \frac{m_\pi^2}{k^2} \ln\left( 1 + 4 \frac{k^2}{m_\pi^2}\right) \bigg)
    \,,
\end{aligned}
\end{equation}
and $k_F(\vec{r}) \equiv [3 \pi^2 \rho (\vec{r})/2]^{1/3}$ is the local Fermi
momentum. 

Although non-zero direct terms arise at higher orders (N$^2$LO, N$^4$LO, and
beyond), the expansion of the direct term does not converge for a realistic pion
mass. The DME for the direct current is essentially an expansion of a
finite-range two-body object in delta functions and their
derivatives~\cite{Dobaczewski2010}, and the range of the pion is too large to
allow the expansion to converge quickly at nuclear density. We therefore
truncate the DME at leading order, in which the finite-range $\bar{c}_3$
contribution to the direct term vanishes completely.  As we will see, however,
this approximation is not bad. 
% and is the same one made in the nuclear matter
% approximation of Refs.~\cite{Menendez2011,Klos2013}.

\section{\label{sec:results}Results}

\subsection{\label{ssec:bgt}Modification of Gamow-Teller strength}

In this section we explore the effects of the two-body axial current on the
$\beta^-$ transition strength function, which reduces to the Gamow-Teller
strength distribution in the absence of two-body currents.  We define a ``bare''
Gamow-Teller operator from Eq.~\eqref{eq:1bc_op_fam} as $O_\text{GT} \equiv
\vec{A}_{1b} /g_A$, and add the two-body current to define a modified
Gamow-Teller operator $O_\text{GT}' \equiv O_\text{GT} + (1/g_A)(
\vec{A}_{2b,\pi} + \vec{A}_{2b, s})$.  The bare Gamow-Teller transition
strength from parent state $\ket{i}$ to daughter state $\ket{f}$ is then $B_{fi} \equiv \lvert \bra{f}
O_{\text{GT}} \ket{i} \rvert^2$, and the full strength is $B'_{fi} \equiv
\lvert \bra{f} O_{\text{GT}}' \ket{i} \rvert^2$. 
The charge-changing FAM computes the linear response to an external field operator and constructs the function $S(\omega)$. This function contains poles at excitation energies of the system with residues equal to the transition strengths. We obtain the derivative of,
e.g., the full transition strength from the full FAM linear response,
$S'(\omega)$, via~\cite{Mustonen2014}
\begin{equation}
\label{eq:dBdE}
    g_A^2 \frac{dB'}{d\omega} = - \frac{1}{\pi} \Im[S'(\omega)]
    \,.
\end{equation}
From the transition strength function we quantify the net two-body effect by
defining the ``total quenching factor'',
\begin{equation}
\label{eq:qft}
q \equiv \sqrt{\frac{\displaystyle \int d\omega\frac{dB'}{d\omega} }{\displaystyle \int d\omega
    \frac{dB}{d\omega}  }}
    \,.
\end{equation}
As the definition shows, $q$ is determined by the ratio of the summed strengths,
and is independent of $g_A$. The total quenching factor allows one to define an
effective value of the axial-vector coupling, $g_A^{\text{eff}} = q\, g_A$, that
could be used in a one-body calculation to account for two-body effects.  In
general, those effects will depend on the transition, causing $q$ to depend on
the energy range of the integrals in Eq.~\eqref{eq:qft}. We therefore consider
QRPA energies up to 60~MeV, which is generally sufficient to exhaust the
$\beta^-$ contribution to the Ikeda sum rule in the nuclei we consider.

To begin our exploration of two-body currents in heavy nuclei we focus on a
small set of nuclei, including the well-studied spherical isotopes $^{48}$Ca,
$^{90}$Zr, and $^{208}$Pb, plus, to examine the effects of neutron excess and
deformation, the spherical isotopic chain $^{132}$Sn--$^{174}$Sn and the
well-deformed isotopic chain $^{162}$Gd--$^{220}$Gd. We include only even-even
isotopes and truncate the chains at the two-neutron drip line.  Finally, to
explore the effects of changes in the total mass, we include the light nuclei
$^{20}$O and $^{28}$O, and the superheavy nuclei $^{294}$Og and $^{388}$Og.  We
find $^{28}$O and $^{388}$Og to be at the two-neutron drip line.

\begin{table}[t!]
\caption{\label{table:LECs}
A summary of the LECs used in this work.  The 
entries all have units of GeV$^{-1}$.  } \renewcommand{\arraystretch}{1.125}
\begin{ruledtabular}
\begin{tabular}{clccc}
   & Label & $c_3$ & $c_4$ & \\
   \hline \noalign{\vskip 1ex} 
   & EGM~\cite{Epelbaum2005} & -3.40 & +3.40 &\\
   & RTD~\cite{Rentmeester2003} & -4.78 & +3.96 &\\
   & EM~\cite{Entem2003} & -3.20 & +5.40 &\\
\end{tabular}
\end{ruledtabular}
\end{table}

\begin{figure}[b!]
\centering
\includegraphics[width=1\columnwidth]{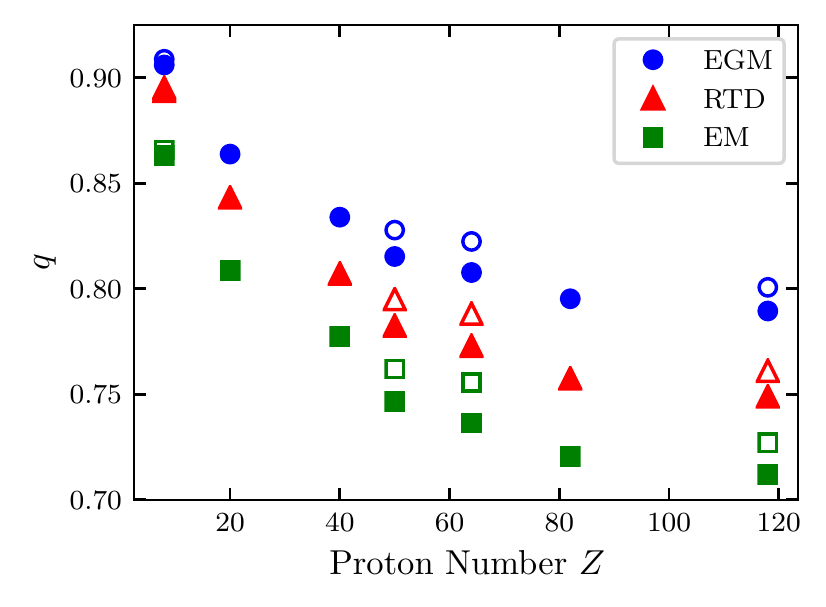}
\caption{\label{fig:qft_vs_z_LECs}%
Total quenching trends with proton number for the different sets of LECs in
Table~\ref{table:LECs}, with $\bar{c}_D=0$. The solid symbols correspond to the
lightest isotope of a given element and the open symbols to the heaviest one at
the drip line.}
\end{figure}

Figure \ref{fig:qft_vs_z_LECs} shows the value of $q$ for all the nuclei in our
data set except those in the middle of the isotopic chains.  Because we use a
density functional with no direct connection to the interactions and operators
of $\chi$EFT, we consider three different sets of LECs for the long-range
contribution to the current (see Table~\ref{table:LECs}) and at first ignore the
contact coefficient $\bar{c}_D$.  We find that the two-body current always has
an overall quenching effect.  For all LEC sets, the amount of quenching
increases with proton number, leading to values of $q$ between 0.86 and 0.91 in
$^{20}$O and between 0.73 and 0.80 in $^{388}$Og.  For elements at the
boundaries of the isotopic chain, we also observe slightly less quenching in the
heavier isotopes than the lighter ones.  The differences in $q$ between $^{20}$O
and $^{28}$O are only $0.0014$--$0.0023$, however, while the differences between
the heaviest and lightest isotopes for Sn, Gd, and Og are all larger and
similar, averaging $0.013$--$0.017$.

\begin{figure}[t!]
\centering
\includegraphics[width=1\columnwidth]{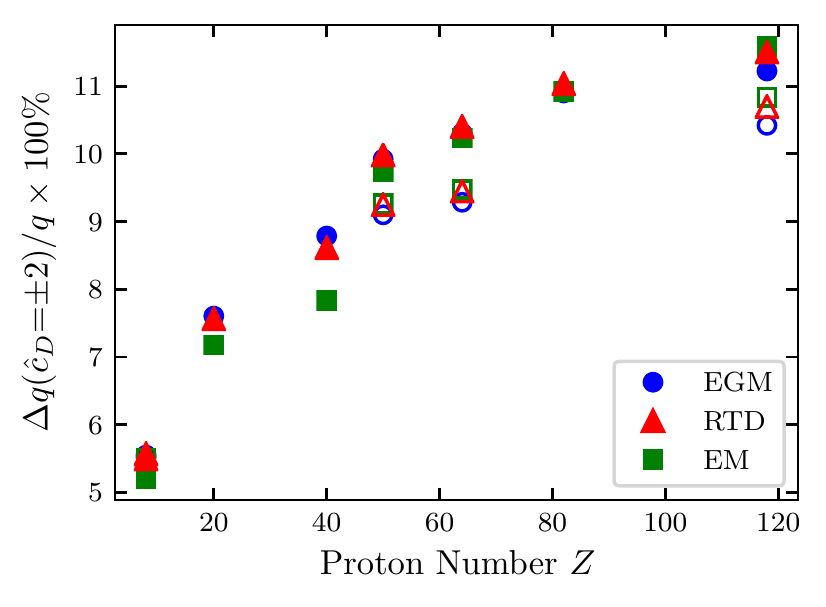}
\caption{\label{fig:qft_vs_z_cD}%
Same as Fig.~\ref{fig:qft_vs_z_LECs} but for
the change in $q$ with $\bar{c}_D = \pm 2$.}
\end{figure}

In Fig.~\ref{fig:qft_vs_z_cD} we examine the effect of the contact term.
Equation \eqref{eq:contact_term} shows that adding a positive $\bar{c}_D$
reduces the quenching while adding a negative $\bar{c}_D$ increases it. The
amount by which $q$ is raised or lowered is almost symmetric about
$\bar{c}_D=0$, so the difference between the $\bar{c}_D= \pm 2$ values in
Fig.~\ref{fig:qft_vs_z_cD} (denoted $\Delta q(\bar{c}_D=\pm2)$) can be thought
of as the size of error bars on the $q$ values in Fig.~\ref{fig:qft_vs_z_LECs}
due to the variation of $\bar{c}_D$ in this range.  The $\bar{c}_D$ contribution
by itself is the same for all LEC sets; small differences in its effects reflect
the interference of the contact with the finite-range term.  The size of the
variation due to $\bar{c}_D$ follows the same trends with $N$ and $Z$ as $q$
itself, and ranges from $5\%$ to $10\%$ of $q$ at $\bar{c}_D=0$.

\begin{figure}[t!]
\centering
\includegraphics[width=1\columnwidth]{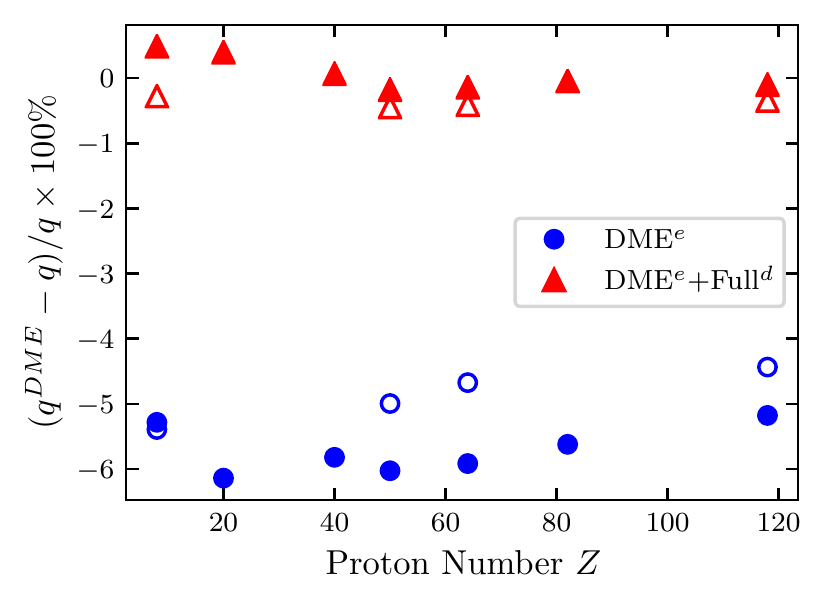}
\caption{\label{fig:qft_vs_z_DME}%
The difference between the total quenching
produced by the density-matrix expansion (DME) and by the full calculation, as a
percentage of the full quenching with the RTD parameter set
\cite{Rentmeester2003}. The blue circles correspond to predictions of the DME
exchange current, not supplemented by anything else, while red triangles correspond
to a supplementation of the DME exchange current by the full direct current (see
text).}
\end{figure}

We compare calculations with the full current to those with the DME current in
Fig.~\ref{fig:qft_vs_z_DME}.  The DME predictions --- the circles in the figure
--- differ from those of the full current by roughly the same amount, $\sim
5\%$, for all nuclei. This indicates that the DME captures the same trends as in
Fig.~\ref{fig:qft_vs_z_LECs}, but for all nuclei it overestimates the quenching
slightly.  The source of this discrepancy is the direct term, which is neglected
in the DME. The squares in the figure add the full current's direct term to the
DME exchange term, producing a small correction that makes the agreement with
the exact results almost perfect. 

\begin{figure}[!hb]
\centering
\includegraphics[width=1\columnwidth]{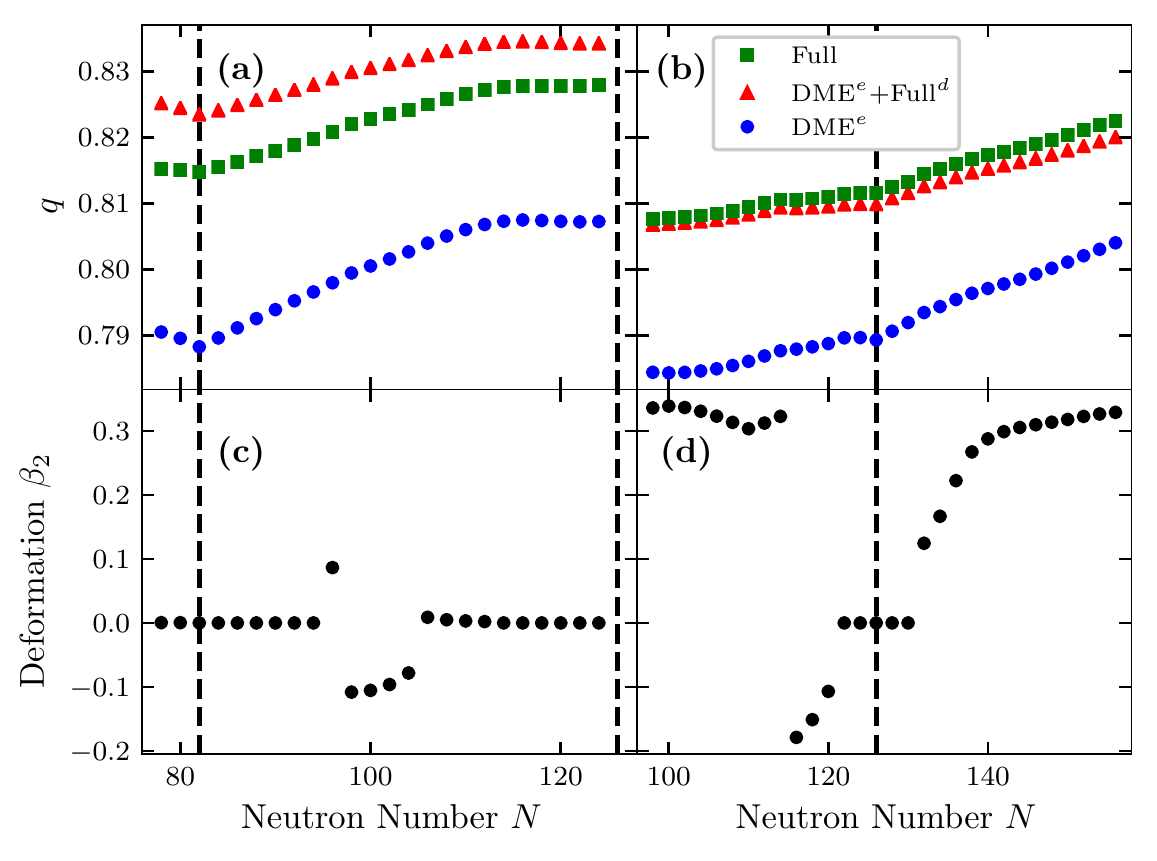}
\caption{\label{fig:Gd_Sn_qtot}%
Total quenching in (a) Sn and (b) Gd isotopic chains using the EGM LEC set with the
full current and the same versions of the DME shown in Fig.\
\ref{fig:qft_vs_z_DME}. (c) and (d) show the respective deformations.
Vertical lines indicate the magic numbers $N=82$ and $126$.}
\end{figure}

Next, we explore trends with deformation and neutron excess by computing the
two-body contributions for all even-even isotopes of Sn and Gd from $N=78$ to
$N=124$ (Sn) and $N=98$ to $N=122$ (Gd). We use the EGM LEC set with
$\bar{c}_D=0$, and again compare the full results with those of the DME.  From
Fig.~\ref{fig:Gd_Sn_qtot}\hyperref[fig:Gd_Sn_qtot]{(c)} we see that the Sn
nuclei are mostly spherical, though those in the middle of the shell are
slightly deformed. On the other hand,
Fig.~\ref{fig:Gd_Sn_qtot}\hyperref[fig:Gd_Sn_qtot]{(d)} shows the Gd isotopes
are very prolate, except right around the shell closure at $N=126$.  Although we
see a small increase in quenching near the shell closures in both elements,
there is no significant trend with deformation.  There is, however, a small,
continuous decrease in quenching with neutron excess that is also apparent in
Fig.~\ref{fig:qft_vs_z_LECs}.  The DME exchange term mirrors the full
calculation, again slightly overestimating the quenching.  We display its
results because they are so much easier to compute than the full results.  Like
the nuclear-matter approximation of Refs.~\cite{Klos2013,Menendez2011}, the DME
exchange expresses the two-body contribution as a density-dependent
renormalization of the one-body Gamow-Teller operator.  We thus expect the
amount quenching to closely mirror the nuclear density. 

\begin{figure*}
\centering
\includegraphics[width=2\columnwidth]{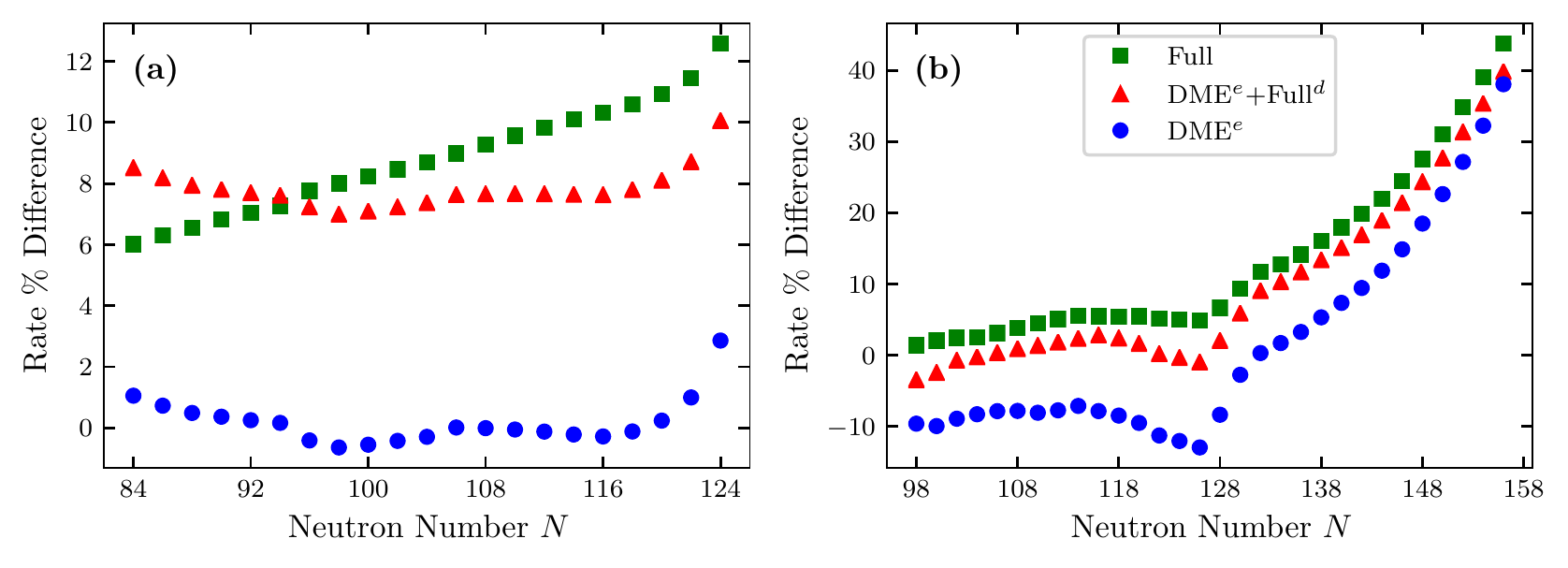}
\caption{\label{fig:Gd_Sn_rates}%
Comparison of Gamow-Teller rates computed with the full
one-plus-two-body current and $g_A=1.27$ to those computed with the one-body
current only and $g_A^{\text{eff}}=1.0$, in (a) Sn and (b) Gd isotopes.
The percent difference is $(\lambda^{\text{1b+2b}}_{g_A=1.27} -
\lambda^{\text{1b}}_{g_A=1.0})/ \lambda^{\text{1b}}_{g_A=1.0} \times 100\%$,
with $\lambda$ representing decay rates. The isotopes $^{128}$Sn--$^{132}$Sn are
excluded because the Gamow-Teller strength below the $\beta$-decay threshold is negligible.}
\end{figure*}

\subsection{\label{ssec:rates}Gamow-Teller rates}

We turn now to $\beta$-decay rates.  They have been measured in only a few of
the nuclei in our set and with the Skyrme functional that we use, we
under-predict those rates even without including a two-body current
\cite{Ney2020}. We are thus not able to see how much the two-body current will
improve the description of rates without re-calibrating the functional, perhaps
even treating the LECs as free parameters, and examining more data.  We leave
that major task for a future publication.  Here, however, we can still get an
idea of what to expect by looking at the differences between rates computed with
the one-body axial current and an effective axial-vector coupling
$g_A^{\text{eff}}=1.0$, employed in most EDF work so far (including
Ref.~\cite{Ney2020}), and those computed with the one-plus-two-body axial
current and the bare axial-vector coupling, $g_A=1.27$.  To what extent does a
nucleus- and energy-independent effective $g_A$ compensate for the omission of
two-body currents? 

We address the question in the Sn and Gd chains in Fig.~\ref{fig:Gd_Sn_rates},
finding that in lighter isotopes the effective axial-vector coupling closely
approximates the two-body current's effect on the rate.  In very neutron-rich
nuclei, however, we begin to see a more significant difference between the two
approaches. The Sn isotopes show a steady increase in the difference with
neutron number, with an uptick near the drip line to about $12\%$.  In the
lighter Gd isotopes, the difference is very small until the $N=126$ shell
closure, after which it increases markedly, to about $40\%$ in $^{220}$Gd.
Although we do not plot the results, we have made the same rate comparison for
the O and Og isotopes in our data set.  In $^{20}$O, the difference between the
quenched one-body and unquenched one-plus-two-body rates is $32\%$ and in
$^{28}$O it is $24\%$, a variation of only about $8\%$.  But in $^{322}$Og, the
difference is $30\%$ and in $^{388}$Og it is $121\%$, a variation of over
$90\%$.  These findings suggest that a constant effective axial-vector coupling
does not adequately account for the effects of two-body currents, particularly
in very neutron-rich nuclei.

To understand the source of the discrepancy in neutron-rich isotopes, we examine
the change in the low-lying one-body Gamow-Teller strength
\textit{distributions} (up to the $\beta$-decay threshold energy) caused by the
two-body current for the lightest and heaviest Sn and Gd isotopes.  This
analysis is not so easy, unfortunately. The FAM requires that the strength
distribution be computed with an artificial Lorentzian width applied to each
transition, but the overlap of the Lorentzian tails from all transitions, in
particular from the Gamow-Teller resonance, prevents us from using the strength
for any one transition to compute the quenching factor for that transition. To
get the best picture, we should use a very small artificial width to minimize
the distortion. The FAM does not converge well close to the real axis, however,
so we use a moderately small half-width of $\gamma = 0.1$~MeV to maintain
sufficient numerical stability.

\begin{figure}[b!]
\centering
\includegraphics[width=1\columnwidth]{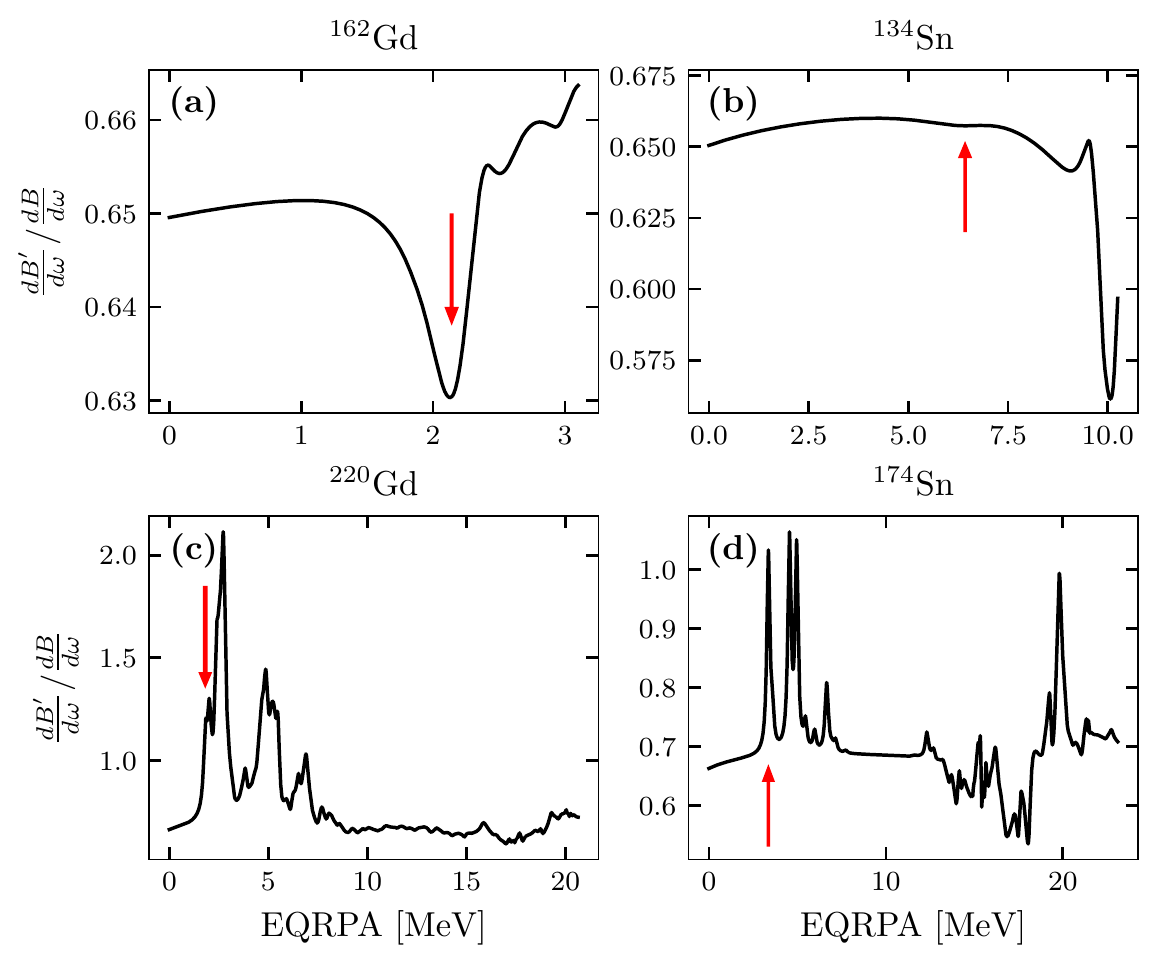}
\caption{\label{fig:Sn_Gd_low_lying_bgt}%
Ratio of the one-plus-two-body FAM strength to the one-body strength for low
Gamow-Teller transition below the $\beta$-decay threshold energy in the light 
isotopes (a) $^{162}$Gd and (b) $^{134}$Sn, as well as the heavy isotopes (c) 
$^{220}$Gd and (d) $^{174}$Sn. Arrows indicate the transitions examined in 
Figs.~\ref{fig:Sn_trans_dens}
and~\ref{fig:Gd_trans_dens}.}
\end{figure}

Figure~\ref{fig:Sn_Gd_low_lying_bgt} shows the resulting ratio of the
one-plus-two-body strength to the one-body strength. The two-body current
appears to affect all low-lying transitions in lighter isotopes in almost the
same way, but in the heavier isotopes it appears to enhance some transitions.
Such enhancement offsets the quenching of other transitions in the computation
of $q$, explaining the significant underestimate made by the effective
axial-vector coupling in these heavier nuclei. 

\begin{figure}[t!]
\centering
\includegraphics[width=1\columnwidth]{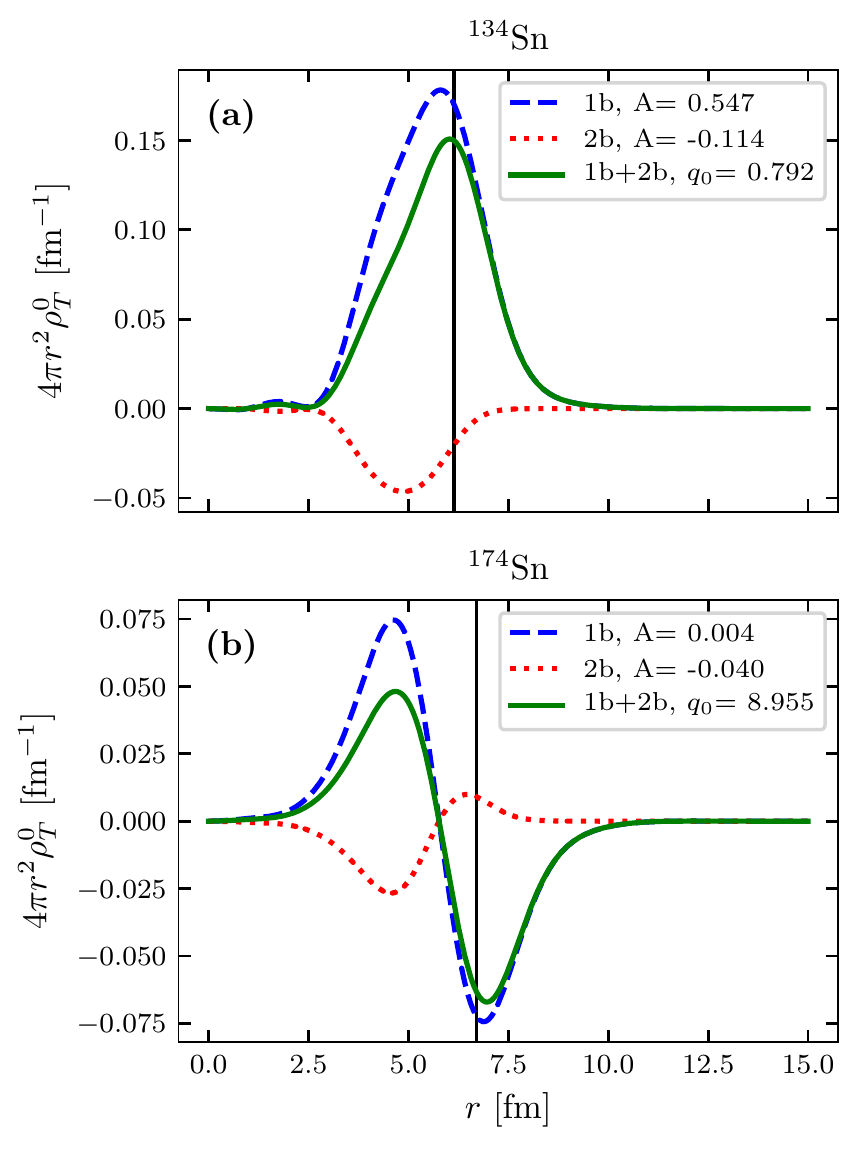}
\caption{\label{fig:Sn_trans_dens}%
Density of the lowest lying Gamow-Teller transition amplitude in (a)
$^{134}$Sn and (b) $^{174}$Sn as a function of the radial coordinate
$r$. The curves show densities for the one-body, two-body, and
one-plus-two-body amplitudes. The factor of $4 \pi r^2$ makes the integrated
amplitudes equal to the areas under the curves. The vertical line is at the
spherical radius $R_0 = 1.2 A^{1/3}$, and the legends show the amplitudes $A$ and
quenching factors $q_0$ for the state in question.  The overall signs of the
amplitudes are arbitrary, but the relative signs between the one- and two-body
terms are not. The quenching factor $q_0$ contains the square root of the
squared amplitudes and is always positive.}
\end{figure}

\begin{figure*}[t]
\centering
\includegraphics[width=.8\linewidth]{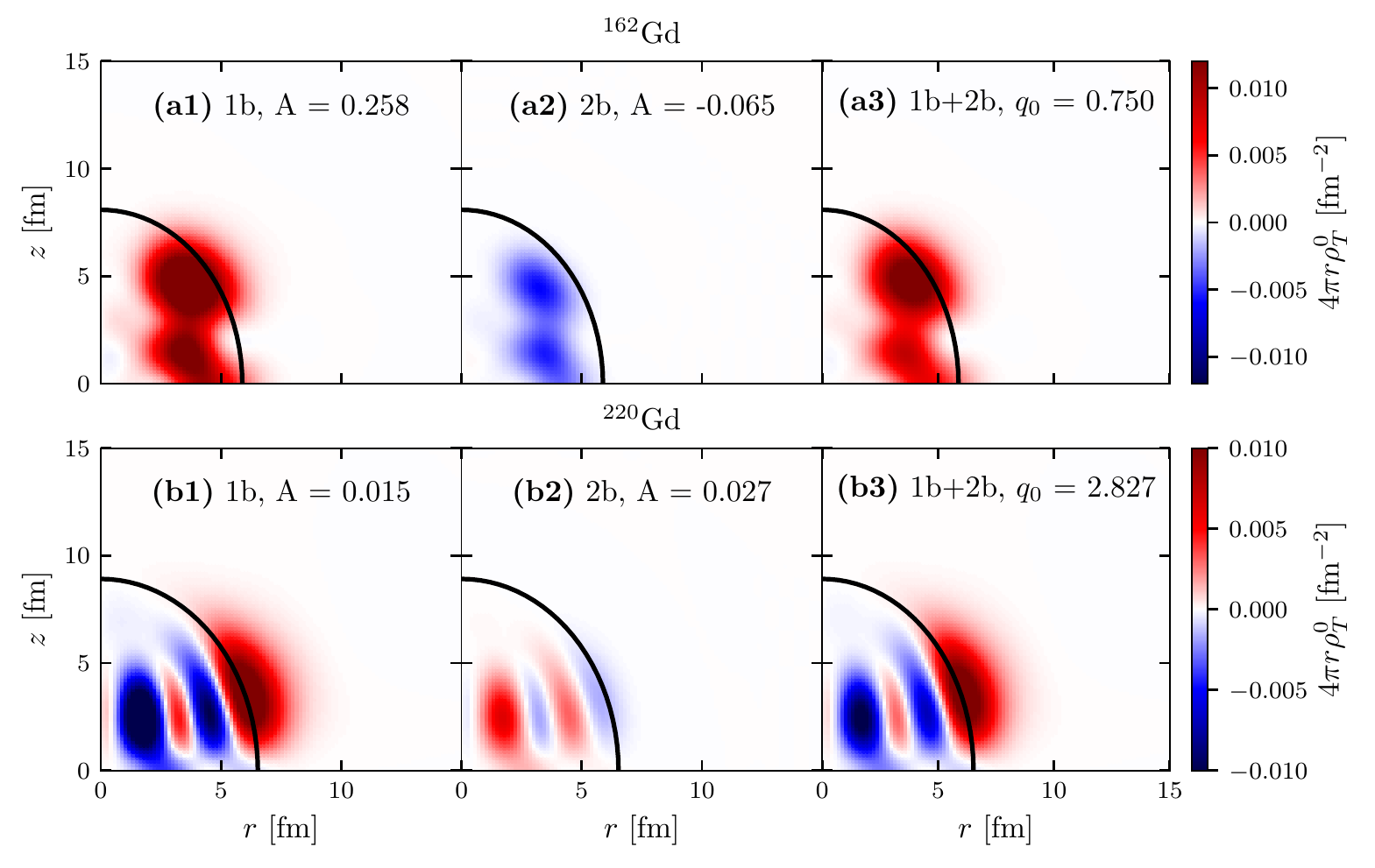}
\caption{\label{fig:Gd_trans_dens}%
Density of the lowest lying Gamow-Teller transition amplitude in (a)
$^{162}$Gd and (b) $^{220}$Gd as a function of $r$ and $z$. The
figures show densities for the (a1,b1) one-body, (a2,b2) two-body, and (a3,b3) one-plus-two-body amplitudes. The volume element of $4 \pi r$ is included (with an extra factor of
$2$ to account for the lower hemisphere). The curved lines indicate the nuclear
surface determined from Eq.~\eqref{eq:spheroid_surface}, with $r = 1.2
A^{1/3}$.  The titles show the amplitudes $A$ and quenching factors $q_0$ for the state.}
\end{figure*}

What causes the enhancement of some low-lying transitions in neutron-rich
isotopes?  To answer, we compute the transition-amplitude density for the lowest
lying states in the heaviest and lightest Sn isotopes with non-negligible rates.
Although the quenching discrepancy in the Sn rates is not as large as in that in
the Gd rates, the Sn isotopes are easier to understand because they are mostly
spherical, allowing us to display the density as a function of a single
variable.  We use the DME to compute the densities because it provides a local
one-body external field, while the field from the full current is non-local.
(Details on the computation of transition densities in the FAM appear in
Appendix~\ref{app:trans_dens}.) Figure~\ref{fig:Sn_trans_dens} shows that the
two-body current contributes very little at or beyond the nuclear surface.  The
reason is the nucleon-density dependence in Eq.~\eqref{eq:DME_exchange} of the
DME current field, which weakens quickly as the density falls.  The fall-off of
the one-body curve is much slower.   When the one-body contribution at the
surface has the same sign as that in the interior, as in
Fig.~\ref{fig:Sn_trans_dens}\hyperref[fig:Sn_trans_dens]{(a)}, this fact is not
important and the two-body contribution quenches the transition amplitude.  But
when it has the opposite sign, leading to relatively small one-body transition
strength, the difference between the quenching effect of the two-body current in
the interior and its almost negligible effect at the surface causes the
integrated matrix element to change sign and have a larger absolute value than
without the two-body contribution, resulting in an enhanced strength.

As we already noted, enhancement is more common in neutron-rich nuclei than in
those closer to stability.  Although a careful and systematic analysis would be
necessary to convincingly identify the physics responsible for the trend, the
presence of a node in the transition-amplitude density associated with the
space-independent operator $\vec{\sigma} t_-$ --- the condition that leads to
enhancement by the two-body current --- implies a mismatch between the shapes of
neutron and proton single-particle wave functions with the same spatial quantum
numbers.  Such a mismatch is much more common in isotopes with a large neutron
excess.

Most of the Gd isotopes are deformed, and an analysis of transition-amplitude
densities is more complicated than in Sn because the density is not constant on
spherical shells.  Nevertheless, we can proceed.  Figure~\ref{fig:Gd_trans_dens}
shows the density for a slice of the upper right quadrant of the nuclei, and we
see that these deformed isotopes exhibit the same phenomena as the Sn isotopes:
quenching when the one-body density has the same sign everywhere and enhancement
when it changes sign, because of a smaller two-body contribution at the surface
than in the interior.  In the heavier isotope here, however, the surface
contribution outweighs the interior contribution even at the one-body level.
The concentration of the two-body contribution, with opposite sign, in the
interior, makes the imbalance even larger and enhances the integrated transition
strength. 

Although in this exploratory paper we are not yet examining the consequences of
the energy- and isospin-dependence of the quenching for total $\beta$-decay
rates, our findings suggest that they will be significant.

\section{\label{sec:conclusions}Conclusions}

We have developed a method to include the contributions of two-body
charge-changing axial currents in the Skyrme-EDF linear-response, and applied
the method to Gamow-Teller strength and $\beta$-decay rates.  From the current,
we construct mean-field-like external-field matrices that can easily be included
in the linear response equations, e.g., through our charge-changing FAM. 
% The contact term reduces to a density-dependent modification of the
% Gamow-Teller operator and can be computed as easily as a one-body external
% field. To treat the finite-range current, we expressed it in terms of
% derivatives acting on Yukawa potentials. Then, by approximating the Yukawa by
% a sum of Gaussians, we leveraged the analytic properties of a Gaussian matrix
% elements to compute the external field in an axially-deformed harmonic
% oscillator basis.
We have also developed a density matrix expansion for the two-body axial
current.  At leading order the expansion reproduces the contact current operator
exactly and replaces the finite-range operator with a density-dependent one-body
Gamow-Teller operator. This approximation reproduces the full linear response
quite well for all the nuclei we studied, and provides a cheap way of including
two-body contributions.  If the direct term is computed exactly, an easier task
in some codes than in our oscillator-based FAM, the expansion works almost
perfectly.
% The finite-range direct contribution explains most of the small discrepancies
% between the two approaches. In the DME, this contribution is zero at leading
% order, and the expansion does not converge for higher orders with a realistic
% pion mass.

To examine the effects on observables, we took the two-body current operators
and the parameters that multiply them from $\chi$EFT.  We found, first, that in
all the nuclei we studied the two-body current quenches the summed Gamow-Teller
strength. The quenching increases significantly with $Z$ and decreases with $N$.
These trends can be understood by the density dependence of the effective
one-body operator produced by the density-matrix expansion.  We also looked at
the energy dependence of the Gamow-Teller strength, finding that the two-body
current causes a nearly constant quenching of decay to low-lying states near
stability but a quenching with significant state dependence and in some cases
even enhancement in very neutron-rich nuclei.  Even though the amount of
quenching of the summed strength changes just a little as $N$ grows, the
enhancement of low-lying strength can cause $\beta$-decay rates to differ
significantly from what would be predicted by a single effective $g_A$.  The
energy dependence in neutron-rich nuclei, like the isospin dependence of the
quenching of summed strength, is connected with nuclear density profiles and the
occurrence of zeroes in the spatial transition-amplitude distribution when the
neutron excess is large.  

Our results open up a number of interesting paths for future projects.  Global
calculations~\cite{Ney2020,Mustonen2016,Marketin2016,Moller2003} indicate that
first-forbidden $\beta$ decay should be important in many nuclei, and our work
should be extended to that channel and then applied to produce global
calculations for $r$-process simulations.  But most important is the marriage of
$\chi$EFT with EDF theory.  We have taken the first step here by including a
chiral current together with a phenomenological density functional, in a way
that is obviously not self consistent.  It would make sense to refit not only
the coupling constants of the functional but also the LECs in the currents.
Once at least some of that is done, better systematic calculations of
$\beta$-decay rates over the entire isotopic chart will become possible.  Here
the DME, which has already been applied to derive EDFs from chiral
potentials~\cite{NavarroPerez2018,NavarroPerez2019} and has been used to obtain
an analogous density-dependent current, will be especially useful.  And with
existing computational technology, one might be able work directly with two-and
three-body chiral interactions and currents, without the DME.  The combination
of EDF phenomenology and methods with \textit{ab initio} interactions and
currents is promising and should be fully investigated. 

\section*{\label{sec:acknowledgments}Acknowledgments}
Thanks to L.J Wang and R. Navarro-Perez for helpful correspondence regarding two-body currents and their numerical implementation.  This work was supported in
part by the Nuclear Computational Low Energy Initiative (NUCLEI) SciDAC-4
project under US Department of Energy Grants No. DE-SC0018223 and No.
DE-SC0018083, and by the Department of Energy under Grant No. DE-SC0013365 and
the FIRE collaboration.  Some of the work was performed under the auspices of
the US Department of Energy by Lawrence Livermore National Laboratory under
Contract No.  DE-AC52-07NA27344. Computing support came from the Lawrence
Livermore National Laboratory (LLNL) Institutional Computing Grand Challenge
program.

% \clearpage
% \newpage
\appendix

\section{\label{app:dme}Density matrix expansion}

Although one might start from a time-dependent energy-density functional that
includes the effects of currents, the result of a DME will be the same as if we
we start with the ``mean-field currents'' $\widetilde{\vec{\Gamma}}_{\pi}$ and $\widetilde{\vec{\Gamma}}_{s}$.
Instead of working with the charge-changing density, as we would in an energy
functional, we (equivalently) obtain the DME exchange functional by applying
Eq.~(24) from Ref.~\cite{Gebremariam2010} directly to the products of
single-particle wave functions (those corresponding to the single particle
states $i$ and $j$ in, e.g., Eq.~\eqref{eq:2b_contact_contract}) as well as to
densities.  Using the 4-component spin-isospin vector $\Psi_{i} (\vec{r})$ in
place of the individual components $\varphi_i (\vec{r}, \sigma, \tau)$, we have
\begin{widetext}
\begin{equation}
\label{eq:dme-basic}
\begin{aligned}
& \Psi_{i}^{\dag} (\vec{r}_1) 
\Psi_{j}(\vec{r}_2 )  
= e^{i \vec{r} \cdot \vec{k}} 
e^{\vec{r}\cdot\left[ \frac{1}{2} (\nabla_1-\nabla_2) -i \vec{k}\right]}
\Psi_i^{\dag} (\vec{r}_1) 
\Psi_j(\vec{r}_2)
\Big|_{\vec{r}_1=\vec{r}_2=\vec{R}} \\
&\simeq e^{i \vec{r} \cdot \vec{k}}  
\left\{1+\vec{r} \cdot\left[ \frac{1}{2} (\nabla_1-\nabla_2) -i \vec{k}\right]
% \right. \\
% & \left. 
+\frac{1}{2}\left( \vec{r}\cdot\left[ \frac{1}{2} (\nabla_1-\nabla_2) -i \vec{k}\right]\right)^{2}\right\} 
\Psi_{i}^{\dag} (\vec{r}_1) 
\Psi_{j}(\vec{r}_2)  
\Big|_{\vec{r}_1=\vec{r}_2=\vec{R}} \,. 
\end{aligned}
\end{equation}
\end{widetext}
Here $\vec{k}$ is a characteristic momentum, $\vec{r} \equiv \vec{r}_1 -
\vec{r}_2$, and $\vec{R} \equiv \tfrac{1}{2}(\vec{r}_1 + \vec{r}_2)$.  

For the decaying nucleon labeled by $i$ and $j$, we take the characteristic
momentum to have magnitude $k_{F}$.  Assuming that the decaying nucleon is near
the Fermi surface, working at the leading-order in Eq.~\eqref{eq:dme-basic} for
the decaying-nucleon states $\ket{i}$ and $\ket{j}$ -- that is, neglecting all
terms in the large braces except ``1,'' -- and averaging over angles for the
momenta $\vec{k}$ (with the assumption that the system is spherically symmetric)
gives 
\begin{equation}
\label{eq:dme-decaying}
\Psi_i^{\dag} (\vec{r}_1) 
\Psi_j(\vec{r}_2)  
\longrightarrow \Pi_1^s\left[ k_F(\vec{R}) r \right] \Psi_i^\dag (\vec{R}) \Psi_j
(\vec{R}) \,, 
\end{equation}
with $\Pi_1^s [kr] \equiv j_0 (kr)$ \cite{Gebremariam2010}.  For the nucleon
densities associated with $\varphi_{k}$ and $\varphi_{l}$ in Eq.\
\eqref{eq:2b_contact_contract}, we use the expansion expressions from Ref.\
\cite{Gebremariam2010}.  If, as for the full current, we neglect the terms in
Eq.~\eqref{eq:2bc_op_fam_y0} that contain $\vec{p}_{1}$ and $\vec{p}_{2}$, the
non-local spin density does not contribute and integrating the chiral current
together with $\Pi_1^s \left[ k_F(\vec{R})r \right]$ and the nucleon density
from Eq.~(26) of Ref.~\cite{Gebremariam2010} over the relative coordinate
$\vec{r}$ eliminates many of the other terms in Eq.~\eqref{eq:2bc_op_fam_y0}.
The integrals, together with the replacement of $\vec{R}$ by the one-body
coordinate $\vec{r}$, result in Eq.~\eqref{eq:2b_contact_exc}.  The expansion
can be continued to higher order in both the wave functions of the decaying
nucleon and the densities associated with the other nucleons, but we do not
present the results of that analysis here. 

The direct part of the current can also be expanded in the manner described in
Ref.~\cite{Dobaczewski2010}, where it was applied to the Gogny interaction.
When used together with the chiral interaction, however, the expansion does not
converge quickly, at least in our tests.  We attribute the problems to the long
range of pion exchange.  In any event, in leading order the direct current does
not contribute at all, so that the most of the effects of the two-body currents
come from the exchange current. 

\section{\label{app:trans_dens}Transition densities}

We wish to understand why the two-body current quenches some low-lying
transitions and enhances others, particularly in neutron-rich nuclei. To do so,
we compute the spatial density of the transition amplitude for particular
transitions.  A transition amplitude to a given state within the FAM depends on
the corresponding QRPA eigenvector, which we must extract.  Once we have it, we
obtain the transition amplitude from 
\begin{equation}\label{eq:QRPA_transition_amp}
\bra{m} F \ket{0} = \sum_{\mu < \nu} \big( X^{m*}_{\mu\nu} F^{20}_{\mu\nu} + Y^{m*}_{\mu\nu} F^{02}_{\mu\nu} \big)
\,,
\end{equation}
where the $X$'s and $Y$'s make up the QRPA eigenvector for state $\ket{m}$.

Reference~\cite{Hinohara2013} explains how the QRPA modes are related to the FAM
response and shows that they can be determined up to an unknown phase
$e^{i\theta} = {\bra{m}\hat{F}\ket{0}}/{\lvert \bra{m}\hat{F}\ket{0}}\rvert$,
from the expression
\begin{equation}
    X^m_{\mu\nu} = e^{i\theta}\ \frac{\text{Res}[X^{\text{FAM}}, \Omega_m]}{\sqrt{\text{Res}[S, \Omega_m]}}
    ,\
    Y^m_{\mu\nu} = e^{i\theta}\ \frac{\text{Res}[Y^{\text{FAM}}, \Omega_m]}{\sqrt{\text{Res}[S, \Omega_m]}}
    \,,
\end{equation}
where $\Omega_m$ is the excitation energy of state $\ket{m}$, $X^{\text{FAM}}$
and $Y^{\text{FAM}}$ are the FAM amplitudes \cite{Avogadro2011}, $S$ is the FAM
response given, e.g., in Eq.~(28) of Ref.~\cite{Hinohara2013}, and
$\text{Res}[A, \Omega_m ]$ is the residue of quantity $A$ at frequency
$\Omega_m$.  One can extract the residues of the FAM quantities from contour
integrals, but it is difficult to choose a contour that contains only a single
transition.  A more efficient but more approximate method to extract the
residues is to compute the FAM quantities a small distance $\gamma$ above the
real axis.  The residue of $S$ is then given by \cite{Litvinova07}
\begin{equation}
\label{eq:approx_s_residue}
\text{Res}[S, \Omega_m]  \approx - \gamma \mathrm{Im}[S(\Omega_m + i\gamma)]
\,.
\end{equation}
A similar relation holds for $X^{\text{FAM}}$ and $Y^{\text{FAM}}$ if we choose
the undetermined phase such that the QRPA eigenvectors are real.  To verify that
the energy at which we are evaluating these functions is indeed a QRPA
eigenvalue, and that the peak is well-separated enough for the approximation in
Eq.~\eqref{eq:approx_s_residue} to hold, we compute the norm of the QRPA mode we
extract:
\begin{equation}
N = \sum_{\mu<\nu} \big( X^{m*}_{\mu\nu} X^m_{\mu\nu} - Y^{m*}_{\mu\nu} Y^m_{\mu\nu} \big)
\,.
\end{equation}
If $N$ is close to one, we can be confident the results are close to the true
QRPA values. The transitions highlighted in Fig.~\ref{fig:Sn_Gd_low_lying_bgt}
all have values of $N>0.99$ for $\gamma=0.01$~MeV.

To obtain a spatial density for the transition amplitude we express
Eq.~\eqref{eq:QRPA_transition_amp} in coordinate space.  Because the mean fields
for one-body and two-body currents are one-body operators, they have the simple form
\begin{equation}
\hat{F} = \sum_{ij} f^{11}_{ij} a^\dagger_i a_j
\,,
\end{equation}
and the (charge-changing) transition-amplitude density can be defined through the
relation 
\begin{equation}\label{eq:trans_dens_def}
\bra{m}\hat{F}\ket{0} = \int d^3r\,\rho_{T}^m(\vec{r})
\,,
\end{equation}
where, for $\beta^-$ transitions,
\begin{equation}\label{eq:trans_dens_dr_f}
\rho_{T}^m(r) = \sum_{\sigma,\sigma'}\int d^3r'\, \delta\rho^m(\vec{r} \sigma p , \vec{r}' \sigma' n) f^{11}(\vec{r}\sigma p , \vec{r}'\sigma'n)
    \,. 
\end{equation}
In Eq.~\eqref{eq:trans_dens_dr_f} the term $\delta \rho^m$ is the density
perturbation for the $m^{\text{th}}$ QRPA mode in coordinate space, obtained
through a change of basis,
\begin{equation}\label{eq:dr_coord}
\begin{aligned}
    &\delta\rho^m(\vec{r} \sigma p, \vec{r}'\sigma' n) 
    % = \bra{i} a^\dagger_{\vec{r}'s 'p} a_{\vec{r}\,s\,n} \ket{0} 
    % \\
    % &= \bra{i} \sum_{ij} \phi^*_j(r's') \phi_i(rs) a^\dagger_j a_i \ket{0}
    % \\&=
    % \sum_{ij} \phi^*_j(r's') \phi_i(rs) \bra{i} a^\dagger_j a_i \ket{0}
    % \\&=
    % \sum_{ij} \phi^*_j(r's') \phi_i(rs)\ \delta\rho_{ij}
    % \\&
    = \sum_{pn} \phi_p^*(\vec{r}'\sigma' p) \phi_n(\vec{r}\sigma n)
    \delta \rho^m_{pn}
    % \\&\qquad
    % \times 
% 
    % \bra{i} \alpha^\dagger_{p_\nu} \alpha_{n_\mu} \ket{0} V_{n_\nu p_j}^*
    \,,
\end{aligned}
\end{equation}
where the $\phi_{k}(\vec{r}\sigma\tau) \equiv \psi^{\abs{\Lambda}}_{n_r n_z}(r,z)
\frac{e^{i\Lambda \phi}}{\sqrt{2\pi}} \chi_{_\Sigma}(\sigma) \chi_{q}(\tau)$ are
axially-deformed oscillator basis states and the $\delta\rho^m_{pn}$ are given by
the inverse Bogoliubov transformation,
\begin{equation}
\delta \rho^m_{pn}=
U_{p\pi} X^m_{\pi \nu} V^{T}_{n \nu} - V^*_{p \pi} Y^m_{\pi \nu} U^{\dagger}_{n \nu}
\,,
\end{equation}
where $U_{p\pi}$ refers to the $p^{\rm th}$ basis component of the 
$\pi^{\rm th}$ proton quasiparticle state and $U^{\dagger}_{n \nu}$ refers to 
the $n^{\rm th}$ basis component of the $\nu^{\rm th}$ neutron quasiparticle 
state.

The full two-body current is non-local in position space, so we use the DME
approximation to get a transition-amplitude density that depends on the local
particle density.  We then define a density-dependent function $g[\rho]$ such
that the one-, two-, and one-plus-two-body current external fields can all be
written in the form
\begin{multline}
\label{eq:f11_coord}
f^{11}(\vec{r}\sigma \tau,\vec{r}'\sigma' \tau') 
= g[\rho(\vec{r})]\ \delta(\vec{r}-\vec{r}') \\
\times\bra{\sigma} \vec{\sigma} \ket{\sigma'} \bra{\tau} t_{-} \ket{\tau'}
\end{multline}
with
\begin{equation}
g[\rho(\vec{r})] = 
\begin{cases}
-1 & \text{one-body} \\
f[\rho(\vec{r})] & \text{two-body} \\
-1 + f[\rho(\vec{r})] & \text{one-plus-two-body}
\,,
\end{cases}
\end{equation}
and $f[\rho]$ the density-dependent function in Eq.~\eqref{eq:DME_exchange}.

Inserting Eqs.~\eqref{eq:dr_coord} and~\eqref{eq:f11_coord} into
Eq.~\eqref{eq:trans_dens_dr_f} we obtain the transition-amplitude density in our
axially-deformed oscillator basis: 
\begin{multline}
\rho_{T}^m(\vec{r}) = g[\rho(\vec{r})]
\sum_{pn} \delta\rho_{pn}^m \psi_p(r,z)\psi_n(r,z)
\\ 
\times \bra{\sigma_p} \vec{\sigma} \ket{\sigma_n}\delta_{\Lambda_p \Lambda_n} 
\,.
\end{multline}
The angular parts of the oscillator wave functions cause terms for which
$\Lambda_p \neq \Lambda_n$ to vanish, so we have replaced them with a Kronecker
delta to make the density real.

Finally, to construct radial plots for the two-dimensional axially-symmetric
density, we can average the density over shells defined by a spherical radius
${r}$ and the deformation $\beta_2$,
\begin{equation}\label{eq:spheroid_surface}
    S(r, \theta) = r \left[ e^{-\frac{1}{2}\beta_2\sqrt{\frac{5}{4\pi}}}\cos(\theta) + e^{+ \beta_2 \sqrt{\frac{5}{4\pi}}}\sin(\theta) \right]
    \,.
\end{equation}
For spherical nuclei $\beta_2=0$, so the surfaces becomes spheres and the value
of the density over the surfaces is constant.

% \clearpage
% \newpage

% \bibliographystyle{apsrev4-1}
% \bibliography{references}

%merlin.mbs apsrev4-1.bst 2010-07-25 4.21a (PWD, AO, DPC) hacked
%Control: key (0)
%Control: author (72) initials jnrlst
%Control: editor formatted (1) identically to author
%Control: production of article title (-1) disabled
%Control: page (0) single
%Control: year (1) truncated
%Control: production of eprint (0) enabled
%

\end{document}